\documentclass[12pt]{iopart}
\usepackage{graphicx}
\usepackage[colorlinks=true,linkcolor=blue,citecolor=red,urlcolor=magenta]{hyperref}
\usepackage{charter} 
\usepackage[top=3cm,bottom=3cm,left=2.1cm,right=2.1cm,a4paper]{geometry}

\begin{document}
	
\title{Swinging small quantum systems out of available values of control parameters}
\author{Xikun Li$^1$ \& Tomasz Sowi\'nski$^2$}

\address{$^1$School of Physics and Optoelectronic Engineering, Anhui University \\ Hefei, Anhui 230601, China}
\address{$^2$Institute of Physics, Polish Academy of Sciences \\ Aleja Lotnik\'ow 32/46, PL-02668 Warsaw, Poland}
\ead{tomasz.sowinski@ifpan.edu.pl}

\begin{abstract}
When a quantum system is prepared in its many-body ground state, it can be adiabatically driven to another ground state by changing its control parameter. However, relying on adiabaticity is experimentally unjustified. Moreover, the target value of the control parameter may occur outside the experimentally accessible range. The indicated target state, however, can still be reached within a clever protocol of temporal changes of the control parameter provided its decomposition into some basis is known. It turns out that such a protocol can be obtained in the framework of the optimal control theory. In this paper, we show how to apply such an optimization scheme to small quantum systems treating interaction strength as the control parameter. We believe that the proposed approach can be creatively extended to various complex quantum systems.  
\end{abstract}
	
\section{Introduction}
The diverse spectrum of exotic phenomena that arise uniquely in many-body physics makes it a compelling platform for applications in quantum simulation, quantum information, and quantum computation~\cite{2018AcinRoadMap,2014NoriRevModPhys,2012BlochNatPhys}. Achieving these goals requires precise control over many-body quantum systems. Especially important is rapid and robust experimental access to the many-body eigenstates corresponding to specific values of control parameters of the Hamiltonian like the interaction strength, the intensity of external confinement, the value of the external field, {\it etc.}~\cite{2014DevoretBook,Glaser2015,2022KochEPJQ}. This possibility would open up a window for studying very exotic phases of quantum matter that are not accessible for usual values of control parameters~\cite{Wen2017,2025GrassRMP}. Of course, from an experimental point of view, adjusting these extreme values of control parameters is a huge challenge or even impossible. It may require very strong electric or magnetic fields, huge laser intensities, or enormous electric currents. Therefore, it would be very vital and promising to have a well-established scheme of reaching extreme-value-parameter ground states by some time-dependent manipulations in a sufficiently small, experimentally accessible, range of control.

In this work, we show that this goal can be achieved using the quantum optimal control techniques, at least for relatively small many-body quantum systems. We leverage the optimal control theory to prepare desired strongly interacting ground states of small quantum systems treating interaction strength as a control parameter that is experimentally bounded to some range of weak interactions. Meanwhile, we employ this technique to speed up the processes of these quantum state preparations and to estimate the minimal time to reach the target state, {\it i.e.}, the quantum speed limit~\cite{Deffner2017}. We select cubic spline functions as the ramp protocol to avoid rapid oscillations and employ the Broyden-Fletcher-Goldfarb-Shanno approach to enhance the fidelity which is the objective function. The optimized ramp protocol is exceptionally efficient and demonstrates no resistance against the impact of control errors. 

Optimal control theory is the state-of-the-art tool~\cite{dalessandro2021,Brif2010} with its application across diverse physical systems, including nuclear magnetic resonance~\cite{KHANEJA2005} and ultracold atoms~\cite{vanFrank2016,Li2018,2019FogartySciPost,2019KahanUniv}. It typically employs two principal classes of optimization algorithms: (i) local optimization strategies like Krotov~\cite{Sklarz2002,krotov1993}, GRAPE~\cite{KHANEJA2005}, CRAB~\cite{Doria2011}, GROUP~\cite{Sorensen2018}, or GOAT~\cite{Machnes2018}; and (ii) global optimization strategies, exemplified by differential evolution~\cite{Storn1997,das2011} or covariance matrix adaptation evolution strategy (CMA-ES)~\cite{Hansen2006}. The method proposed by us in this work belongs to the first class since it relies on local derivatives. Although analytic solutions are available for quantum systems with low-dimensional Hilbert space~\cite{Lloyd2014,Boscain2006,Hegerfeldt2013,Jafarizadeh2020}, the high-dimensional quantum systems require the invocation of numerical optimization techniques.

The paper is organized as follows. In Sec.~\ref{sec:model} we describe the theoretical framework and the method that forms the basis of our investigation. Then in Sec.~\ref{sec:twospin} we introduce a two-qubit toy model to illustrate the optimization method which is employed in Sec.~\ref{sec:1k2li} to discuss a more realistic model of three interacting fermions. Finally, in Sec.~\ref{sec:threecomponent} we extend the discussion to a larger system of three-component fermionic mixture. In this way, we can demonstrate how the method can be generalized to cases when the optimization is required only for a selected subsystem. Section~\ref{sec:conclusion} concludes our work.

\section{The framework and the method}\label{sec:model}
In our work, we assume that a specific quantum system is described by the time-dependent Hamiltonian of the form
\begin{equation} \label{Hamiltonian}
	{\cal H}(g(t)) = {\cal H}_0 + g(t) {\cal H}_c,
\end{equation}
where ${\cal H}_0$ and ${\cal H}_c$ are noncommuting time-independent parts (the drift and the control Hamiltonian, respectively) and a whole time-dependence comes from the external control field $g(t)$, {\it e.g.}, interaction strength. We assume that the intensity of this field can be quite well controlled experimentally. Of course, this description also includes time-independent scenarios with particularly chosen intensities of the field, $g(t)=g$. In these cases, at least in principle, one can solve eigenproblem for a corresponding Hamiltonian
\begin{equation}
{\cal H}(g)|\Psi_i(g)\rangle = E_{i}(g)|\Psi_i(g)\rangle
\end{equation}
and eigenstates obtained for different intensities $g$ are connected by adiabatic varying of the field. This is particularly true for the ground states $|\Psi_0(g)\rangle$ which remain isolated for any $g$. Already at this level it is interesting to ask the question if it is possible to engineer time evolution of the intensity $g(t)$ such that two different ground states, {\it i.e.}, the initial $|\mathtt{ini}\rangle=|\Psi_0(g_1)\rangle$ and the target $|\mathtt{tar}\rangle=|\Psi_0(g_2)\rangle$ state, are connected (as fast as possible) by unitary finite-time evolution determined by ${\cal H}(g(t))$. Formally this question can be formulated as an optimization problem for finding the intensity $g(t)$ maximizing the final fidelity
\begin{equation} \label{TempFid}
	F(g(t),T)=|\langle\mathtt{tar}|{\cal T}\!\exp\left[-\frac{i}{\hbar}\int_0^T  {\cal H}(g(t)) \mathrm{d}t
	\right] |\mathtt{ini}\rangle|^2,
\end{equation}
where $\mathcal{T}$ is the time-ordering operator and $T$ is the total duration. In our work, to calculate resulting fidelity for a given $g(t)$, we perform the time evolution of the state of the system by solving the Schr\"odinger equation written in the basis of Fock states using the MATLAB function \texttt{ode45} which is based on the Runge-Kutta method.

At this point, it is important to mention that physically this kind of optimization is not sufficiently well-defined since it still does not take into account experimental limitations on control intensity $g(t)$. For example, although it is mathematically possible, in practice, it is not feasible to change the intensity arbitrarily fast. Typically, its amplitude is also limited to some well-defined, experimentally accessible range. Therefore, we should consider these limitations when constructing the control protocol. In the following, we assume that the intensity $g(t)$ can be easily changed only in some range $g\in [g_A,g_B]$. From the physical engineering point of view, the most interesting cases are of course those in which target interaction $g_2$ is essentially outside the range $[g_A,g_B]$, {\it i.e.}, when the target state {\it cannot} be attained with any adiabatic-like protocol. To check these scenarios we consider three substantially different accessible ranges: {\it (i)} $g_2>g_B=-g_A$; {\it (ii)} $g_A=0$, $g_B<g_2$; {\it (iii)} $-g_2<g_A$, $g_B=0$.
Additionally, to avoid nonphysical rapid changes of the control parameter, we introduce an \textit{additional} parameter $M$ encoding the number of equally spaced time points in period $T$ at which the value of $g(t)$ is optimized. Between these points, the value of $g(t)$ is interpolated smoothly via cubic splines (standard \texttt{interp1} function in MATLAB).

At this point it is worth mentioning that without loss of generality, we can assume that initially the system is prepared as the ground state of the drift Hamiltonian ${\cal H}_0$ ($g_1=0$). This is clear given that the division of the Hamiltonian (\ref{Hamiltonian}) to the drift and control part is not unique. Suppose the initial state is the ground state of the Hamiltonian for some nonzero $g_1$. In that case, one can straightforwardly define the drift Hamiltonian as $\tilde{\cal H}_0={\cal H}_0 + g_1{\cal H}_c$ and shift the control field to $\tilde{g}(t) = g(t)-g_1$. Since the performance of the optimization algorithm is insensitive to this change, the optimal solution can be found for any initial $g_1$.

After setting up the objective function and the control protocol $g(t)$, we now employ the optimal control theory to optimize the fidelity $F$, to estimate the quantum speed limit $T_{\mathrm{QSL}}$, and to obtain the optimal control field $g(t)$. We wish to determine the temporal shape of $g(t)$ for which the final fidelity $F(g(t),T)$, for a chosen physical limitation established by $g_A$ and $g_B$, is saturated as close to $1$ as possible. To achieve this goal, we optimize $g(t)$ to obtain the maximal possible fidelity $F(g(t),T)$ for a given $T$. Given $T$, we gradually increase $M$ and repeat the optimization. In general, the fidelity obtained converges upon increasing $M$. In such a case, the optimization for the given value of $T$ is stopped. Then we increase $T$ and do the optimization with this given $T$. We repeat this procedure until fidelity is saturated above a given threshold. The choice of the threshold value $F$ depends on the quantum system to be optimized, as well as the precision wished to obtain. As the precision increases, in general, the total duration also has to be increased. The minimal duration obtained numerically is a good approximation of the quantum speed limit $T_{\mathrm{QSL}}$. In our approach, we choose the Broyden-Fletcher-Goldfarb-Shanno method as the optimization algorithm~\cite{chong2013}. This method can be understood as a quasi-Newton method for optimizing functions that have continuous first and second derivatives. It approximates the Hessian matrix inverse and finds a local minimum of function iteratively.

\section{Two-qubit toy model}\label{sec:twospin}
Before going to more complicated many-body systems, let us first start with an illustration of the method's performance on quite a simple two-qubit model. Let us consider the system described by the Hamiltonian (\ref{Hamiltonian}) with
\numparts \label{HamSigma}
\begin{eqnarray}
	{\cal H}_0 &= -\frac{\Delta}{2\sqrt{2}}\left(\sigma_1^x + \sigma_2^x + \sigma_1^z + \sigma_2^z\right), \\
    {\cal H}_c &= -\Delta\sigma_1^z \sigma_2^z,
\end{eqnarray}
\endnumparts
where $\sigma^x_i$ and $\sigma^z_i$ are Pauli matrices acting on $i$-th qubit. By convention, we introduced numerical coefficients such that the energy splitting between two states of qubit (the difference between eigenvalues of the drift Hamiltonian ${\cal H}_0$) is equal $\Delta$, while the intensity $g(t)$ is a dimensionless parameter controlling the interaction strength between qubits. Consequently, in this model time is naturally measured in units of $\hbar/\Delta$. This model generalizes the exactly solvable two-level system and, although the Hamiltonian can be easily diagonalized for any temporal value of $g$, the closed-form analytical solution for the state preparation problem is unknown. The system is known to have an unexpectedly rich control phase diagram~\cite{Bukov2018}. It can be viewed also as a simplified model for the control problem of coupled quantum oscillators~\cite{2018LiewPRB,2018StefanatosPRB,2020StefanatosPRA}.

Suppose that our goal is to prepare the target state $|\mathtt{tar}\rangle$ which is the ground state of the Hamiltonian for $g_2=4$ by starting from the initial state $|\mathtt{ini}\rangle$ being the non-interacting ground state ($g_1=0$). Moreover, we assume that we have experimental access only to interactions bounded by $g_B=-g_A=2$, which means that the target state cannot be obtained by a direct adiabatic protocol. We aim to reach the state $|\mathtt{tar}\rangle$ by optimizing time-dependence $g(t)$ to maximize the final fidelity $F$ after fixed protocol time $T$ with the method described in the previous section. 
\begin{figure}
\includegraphics[width=\linewidth]{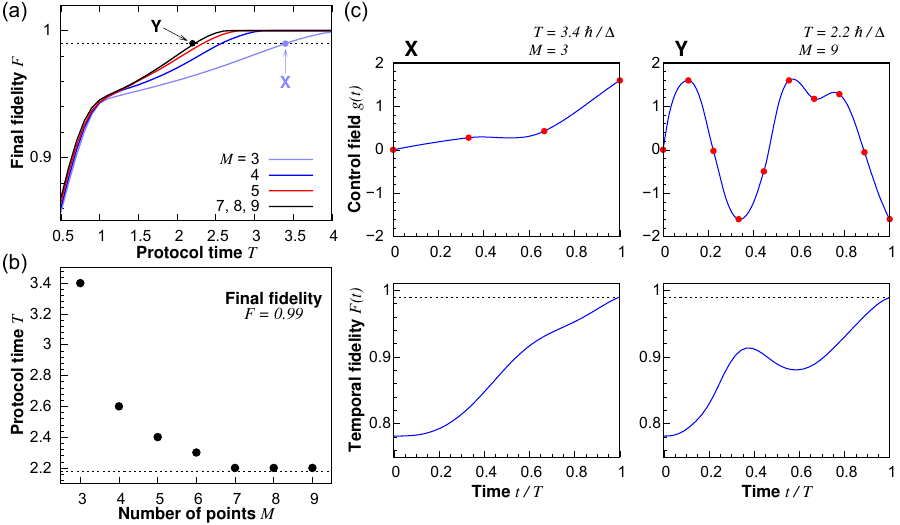}
\caption{Optimization scheme for a two-qubit model with experimental limitation of allowed interaction strengths $g\in[-2,2]$. (a) The final fidelity $F$ as a function of the protocol period $T$ obtained for different numbers $M$ of optimization instants. Along with increasing $M$, the final fidelity saturates at maximally allowed value. Points {\bf X} and {\bf Y} mark protocol time $T$ for which the final fidelity approaches $99\%$ for $M=3$ and $M=9$ respectively. (b)  The minimal duration T needed to reach the final fidelity $F = 0.99$ for a given number of points $M$. (c) Two examples of an optimized time sequence of interactions $g(t)$ leading at finite time $T$ to the target state $|\mathtt{tar}\rangle$  with fidelity $F=99\%$. These examples correspond to points {\bf X} and {\bf Y} marked in plot (a). In plots (a) and (b) time is expressed in natural units $\hbar/\Delta$. 
	}\label{Fig1}
\end{figure}

First, let us note that the fidelity between the initial and the target state is $|\langle\mathtt{tar}|\mathtt{ini}\rangle|^2 \approx 0.7815$. Therefore, this fidelity is a trivial lower bound for optimization since it can be obtained by keeping strength constant in time,  $g(t)\equiv 0$. It corresponds to $M=0$ optimization parameters. By increasing $M$ and performing optimization one can increase the final fidelity. In Fig.~\ref{Fig1}a we present the dependence of the highest possible final fidelity $F(T)$ obtained for a different number of optimization points $M$ and required protocol time $T$. It is clear that for a fixed period $T$, one can increase the maximal fidelity which eventually saturates at $1$. However, for shorter periods $T$, reaching the fidelity close to $1$ is not possible even for a large number of optimization points $M$. For example, the final fidelity larger than 99\% (dashed line) cannot be obtained for periods smaller than limiting period $T_L\approx 2.2 \hbar/\Delta$ since a further increase of the number of points $M$ does not change the performance (black curve). Complementarily, this fact can be deduced from Fig.~\ref{Fig1}b presenting the minimal duration $T$ needed to reach the final fidelity $F=0.99$ for a given number of points $M$. It is clear that for periods shorter than $T_{L}$ (dashed line), an arbitrarily large number of points $M$ cannot guarantee saturation of the final fidelity over $99\%$. This is a direct manifestation of the well-known fundamental limit of non-adiabatic protocols -- the quantum speed limit~\cite{Frey2016}. For completeness of the discussion, in Fig.~\ref{Fig1}c we present two exemplary solutions of the optimization procedure performed for two different situations (defined by the number of points $M$ and the protocol period $T$). In the top row, we present the time dependence of the control field $g(t)$ leading to the final fidelity on the level of $99\%$. The bottom row presents corresponding temporal fidelity with the target state $F(t)$. These two examples correspond to two points marked in Fig.~\ref{Fig1}a. 

\begin{figure}
	\includegraphics[width=\linewidth]{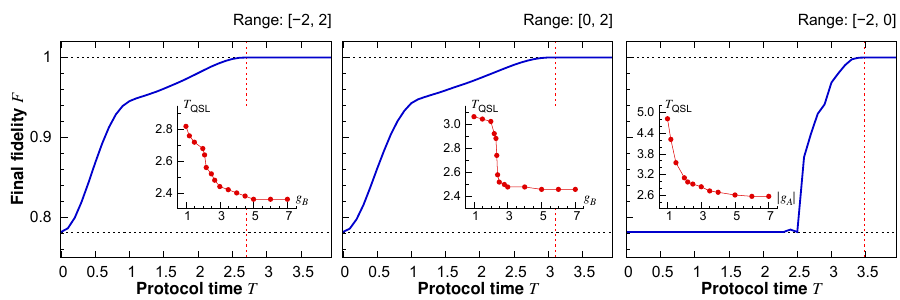}
	\caption{Maximal possible final fidelity $F$ as a function of the total protocol time $T$ for three different experimental limitations set on external strength $g$. The lowest bound of the fidelity is presented by the bottom horizontal dashed line. The vertical dashed red line indicates the estimated quantum speed limit time $T_{\mathrm{QSL}}$. Insets present the estimated $T_{\mathrm{QSL}}$ when the limitation of the range (depending on the scheme) is enlarged. See main text for details. In all plots, time is measured in units $\hbar/\Delta$.}\label{Fig2}
\end{figure}
Similar analysis can be performed for two other experimental limitations, {\it i.e.}, when we have access only to positive or negative values of the control parameter, $g(t)\in [0, 2]$ or $g(t)\in [-2,0]$, respectively. In Fig.~\ref{Fig2} we compare the results for all three scenarios with the same initial and target states. Plots show the maximal possible fidelity $F_\mathrm{max}$ as a function of total duration $T$ obtained by optimization over the number of points $M$. The vertical red dashed line indicates the estimated quantum speed limit time $T_{\mathrm{QSL}}$, {\it i.e.}, the minimal period $T$ for which the optimized fidelity $F_\mathrm{max}$ can be saturated close to $1$ (in the case studied, the fidelity can be saturated at $F=1- \delta$, with $\delta<10^{-10}$). It can be viewed as the limit of previously discussed limiting time $T_L$ when the desired final fidelity approaches $1$. It is clear that for all three scenarios obtaining the target state $|\mathtt{tar}\rangle$ with high fidelity is possible. However, stronger limitations put on the range of accessible values of the control parameter lead to an evident increase in the minimum propagation time. In particular, when the requested target state is defined for an interaction of opposite sign, a meaningful improvement of the final fidelity from its lower bound requires a significantly long protocol. However, the simple fact that it is possible to arrive at a strongly repulsive ground state using only weak non-positive interactions is appealing. Similar conclusions can be obtained for arbitrary values of $g_2$ and the range $[g_A, g_B]$. In general, for a given $g_1$ and fixed range $[g_A, g_B]$ a limiting $T_{\mathrm{QSL}}$ increases with $g_2$. 

In our approach, the \textit{quantum speed limit} refers to the minimal time required to transfer the system from the initial state $|\mathtt{ini}\rangle$ to the target state $|\mathtt{tar}\rangle$ with constrained control field. Thus, it is natural to check the dependence of the resulting $T_{\mathrm{QSL}}$ on the tightness of these constraints. For this purpose, we repeated calculations for all three scenarios after releasing the accessible range of the control amplitude $g$, {\it i.e.}, we enlarge the available interval in the three scenarios accordingly as $[-g_B,g_B]$, $[0,g_B]$, and $[g_A,0]$. Corresponding $T_{\mathrm{QSL}}$ obtained in this way are present in insets in Fig.~\ref{Fig2}. It is clear that as the range becomes wider, the estimated quantum speed limit time $T_{\mathrm{QSL}}$ decreases, and eventually it saturates to a threshold value. In principle, the saturation value of $T_{\mathrm{QSL}}$ can be understood as the global unconstrained lower bound for a given type of protocol. Detailed analysis of this bound may be very interesting and fruitful from the theoretical and fundamental perspective~\cite{2009CanevaPRL,2022HornedalNJP,2024VikramPRL}, however, it is far beyond this work.

Finally, let us also mention that in principle the fidelity itself being close to $1$ does not necessarily mean that the target state of the system $|\mathtt{tar}\rangle$ is well reproduced at the end of the protocol. The final state may contain a small but significant admixture of eigenstates with significantly different energies which would result in large nonequilibrium oscillations in further dynamics. To exclude this possibility we compare expectation values of the final Hamiltonian in the target state and the state at the final stage of evolution. We find, that in all the cases studied the difference quickly vanishes when fidelity approaches $1$. For example, in the case of a symmetric range of accessible interactions (the first scenario), the target energy $E_{\mathtt{tar}}/\Delta=\langle H(g_2) \rangle_{\mathtt{tar}}/\Delta\approx-4.7363$, while the energy of the final state with fidelity $F=99\%$ $E_\mathtt{fin}/\Delta=\langle H(g_2) \rangle_T/\Delta\approx-4.7092$. For the final state having almost perfect fidelity with the target state, $F=1-{\cal O}(10^{-13})$, the difference between these energies is of the order of ${\cal O}(10^{-13})$.

\section{Mixture of three fermions}\label{sec:1k2li}
To illustrate the method for quantum systems having larger Hilbert spaces, let us now consider a two-component mixture of three interacting ultracold fermions confined in a one-dimensional harmonic trap of a given frequency $\omega$. We assume the simplest possible scenario in which fermions belonging to different components interact only via zero-range forces while intra-component interactions are not present. The Hamiltonian (\ref{Hamiltonian}) of this system  reads:
\begin{numparts} 
\begin{eqnarray}\label{eq:hamiltonian}
\hat{\cal H}_0 &= - \frac{\hbar^2}{2 m_A} \frac{\partial^2}{\partial x^2} + \frac{m_A\omega^2}{2} x^2 +\sum_{i=1}^{2} \left[ - \frac{\hbar^2}{2m_B}
		\frac{\partial^2}{\partial y_i^2} + \frac{m_B\omega^2}{2} y_i^2 \right], \label{Ham2-1}\\
\hat{\cal H}_c &= \sqrt{\frac{\hbar^3\omega}{m_B}}\sum_{i=1}^{2}\delta(x - y_i), \label{Ham2-2}
\end{eqnarray} 
\end{numparts}
where $m_A$ and $m_B$ are masses of particles from different components. For convenience, we introduced an additional scaling factor in the control Hamiltonian (\ref{Ham2-2}) assuring that control parameter $g(t)$ is dimensionless. In the following, we focus on the experimentally relevant scenario of $m_A/m_B=40/6$ corresponding to the $^{40}$K-$^6$Li mixture. However, all the results can be straightforwardly generalized to other desired mass ratios similarly.

Properties of different systems described with generic Hamiltonian (5) are deeply studied in many different contexts~\cite{2012BlumeRPP,2019SowinskiRPP,2023MistakidisPhysRep}. It is known that it can be relatively easily diagonalized numerically for any interaction strength $g$. In our approach, we use straightforward diagonalization in the Fock basis spanned by a set of the lowest single-particle orbitals of the harmonic oscillator cut on some, sufficiently large excitation ${C}$. The cut-off is determined operationally by checking the convergence of the final results after a further increase of the basis. We find that all our results presented in the following are well-converged if the cut-off is ${C}=14$. With our method, we are able to determine numerically the lowest eigenstates of the Hamiltonian $|\Psi_i(g)\rangle$ and their eigenenergies $E(g)$. It is also possible to perform time propagation of any state $|\Psi(t)\rangle$ via the Runge-Kutta method, provided that during the evolution one can neglect couplings to cut off Fock states. 
\begin{figure}
	\includegraphics[width=\linewidth]{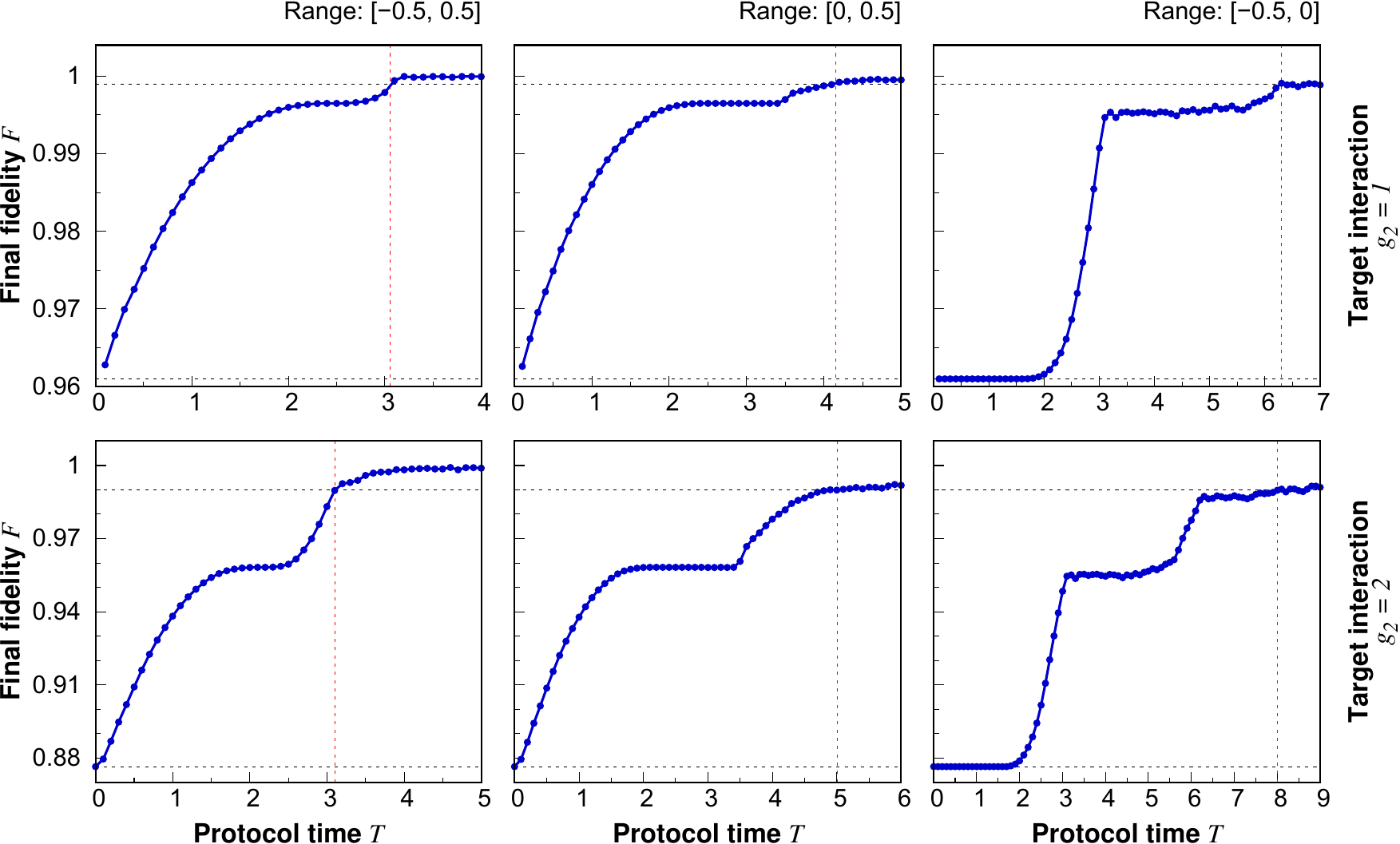}
	\caption{Maximal possible final fidelity $F$ for the system of three fermions as a function of the total protocol time $T$ in the case of three different experimental limitations on external strength $g$ and for two different target interactions $g_2$. The lowest bound of the fidelity is presented by the bottom horizontal dashed line. The vertical dashed red line indicates the estimated quantum speed limit time $T_{\mathrm{QSL}}$ for obtaining fidelity on the level of $99.9\%$ and $99\%$ (respectively for $g_2=1$ and $g_2=2$) indicated by an upper horizontal dashed line. In all plots, time is measured in natural time-unit of harmonic oscillator $1/\omega$.}\label{Fig3}
\end{figure}

It is also interesting to note that specific flattenings of the final fidelity are clearly visible in the vicinity of protocol time $\pi/\omega$. This may suggest that they are somehow related to the trap frequency. Unfortunately, it is also clear that flattenings' exact position and width are essentially and non-trivially dependent on the target interaction $g_2$ and the allowed range of interactions $[g_A,g_B]$. A more thorough explanation of their nature would therefore require further numerical and analytical investigation, which is beyond the scope of this work.

Let us assume that initially the system is prepared in the non-interacting ground state of the system $|\mathtt{ini}\rangle=|\Psi_0(0)\rangle$ and we aim to reach the target state $|\mathtt{tar}\rangle=|\Psi_0(g_2)\rangle$  in finite time $T$ having experimental access only to strengths bounded by $|g|\leq 0.5$. As anticipated in Sec. 2, we consider three different scenarios with $g\in [-0.5,0.5]$, $g\in [0,0.5]$, and $g\in [-0.5,0]$. Determination of the maximal possible final fidelity for this system is straightforward. First, we optimize the time-dependence of the interaction strength $g(t)$ for a given protocol period $T$ and a fixed number of optimization instants $M$. Then, we increase the number of points $M$ and we monitor a saturation of the fidelity to its upper bound. In this way, we obtain the maximal possible fidelity which is presented in Fig.~\ref{Fig3} for three different experimental limitations and two different target interactions, $g_2=1$ and $g_2=2$. In all the cases the final fidelity can be saturated close to $1$ for sufficiently large protocol time $T$ and, of course, for smaller $g_2$ saturation becomes faster. Moreover, similarly, as in the case of a two-qubit system, access to a wider range of interactions supports the reduction of the minimal time required. This fact is well-reflected by estimated quantum speed limit time $T_\mathrm{QSL}$ which strongly depends on the assumed range. Also in this case, for the opposite-sign interactions scheme, the final fidelity is resistant to change if the protocol time is not sufficiently long. 

\begin{figure}
	\includegraphics[width=\linewidth]{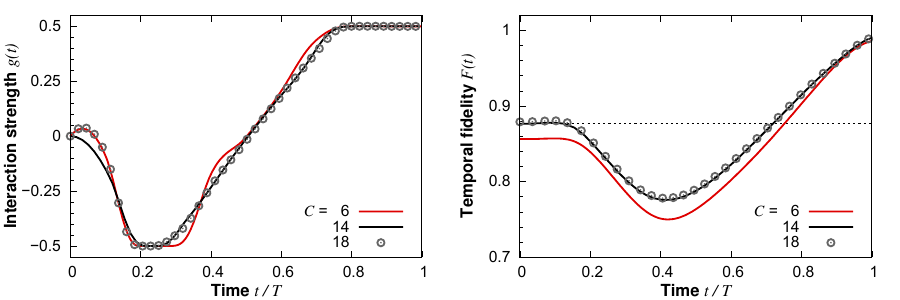}
	\caption{Time evolution of interaction strength $g(t)$ and corresponding temporal fidelity $F(t)$ for the system of three fermions, obtained for different numbers of single-particle orbitals $C$. We notice that for $C=14$ further increasing of the cut-off does not change the results and they are well converged. In both plots the system parameters are fixed at $T=3.5\omega^{-1}$, $g_2=2$ and $-g_A=g_B=0.5$}.\label{Fig4}
\end{figure}
To show that our results are well-converged in terms of the assumed cut-off ${C}=14$ of single-particle basis, in Fig.~\ref{Fig4} we present example results of the optimization procedure,  with the target interaction $g_2=2$,  performed for fixed protocol time $T=3.5\omega^{-1}$ and the number of optimization instants $M=10$ obtained under different cut-offs ${C}$ assumed. When the assumed cut-off is insufficient (red solid line) the target state $|\mathtt{tar}\rangle$ is not determined appropriately and even initial fidelity is not accurate. However, the optimization scheme enables one to find interaction path $g(t)$ saturating the final fidelity close to $1$. Along with increasing cut-off ${C}$ optimization is improved and for a sufficiently large value its further increase does not change the results (black solid line and black circles). 

\begin{figure}
	\includegraphics[width=\linewidth]{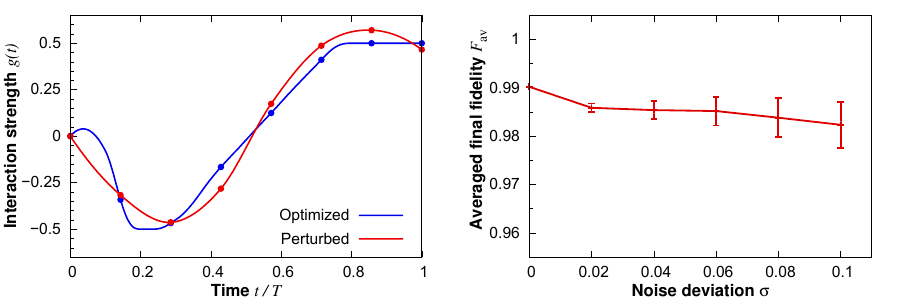}
	\caption{(left) Exemplary realization of the noisy control field (red) obtained from the ideally optimized field (blue) by application of random shifts of optimized values of interaction $g(t)$ at optimization instants (blue and red dots) with noise deviation $\sigma=0.06$. (right) The averaged final fidelity as a function of noise deviation $\sigma$ obtained in the ensemble of 100 randomly generated noisy control fields. Error bars correspond to the standard deviation of the average. In both plots, the system parameters are fixed as $T=3.5\omega^{-1}$, $g_2=2$, and $-g_A=g_B=0.5$.
	}\label{Fig5}
\end{figure}

Finally, we also checked the resistance of the protocol to experimental imperfections in controlling the field. For this purpose, we modeled such imperfections in the simplest possible way, by introducing random perturbations to the values of the already optimized control field $g(t)$ at optimization instants defined by $M$. These perturbations are drawn from a normal distribution centered at zero and some standard deviation $\sigma$. In this way, we consider a randomly perturbed control field and we check its consequences on the final fidelity. An illustration of the noisy control field (with noise deviation fixed $\sigma=0.06$) is provided in the left panel of Fig.~\ref{Fig5}. A single realization of the noisy control field (red line) is established from the optimized control field (blue line) obtained for the case studied previously, {\it i.e.}, $T=3.5\omega^{-1}$, $g_2=2$, and $-g_A=g_B=0.5$. Note that perturbation of the optimized control field $g(t)$ may significantly change a resulting field in a whole considered domain. For such a noisy control field we calculate the final fidelity $F$ and, to make the analysis more meaningful, we repeat this construction 100 times and average. The averaged final fidelity $F_{\mathrm{av}}$ obtained in this way for different noise deviations $\sigma$ is presented in the right panel of Fig.~\ref{Fig5}b. As suspected, an introduction noise introduced to the control field diminishes the quality of the performance of the protocol. However, even for evidently strong randomness in the system, the protocol is quite robust and obtained final fidelity may be considered as satisfied. 

\section{Interactions with third system}\label{sec:threecomponent}
Finally, we analyze the minimal extension of the problem to a situation in which the system of interest is affected by the external surroundings due to uncontrolled interactions. For this purpose, we consider the previous system of three fermions interacting additionally (via zero-range forces) with a third-component particle of mass $m_B$ confined in the same harmonic potential. The Hamiltonian of the system considered, when written as (\ref{Hamiltonian}), reads:
\begin{numparts} \label{hamiltonian3}
\begin{eqnarray}
\hat{\cal H}_0 &= - \frac{\hbar^2}{2 m_A} \frac{\partial^2}{\partial x^2} + \frac{m_A\omega^2}{2} x^2 +\sum_{i=1}^{2} \left[ - \frac{\hbar^2}{2m_B}
		\frac{\partial^2}{\partial y_i^2} + \frac{m_B\omega^2}{2} y_i^2 \right]+ \nonumber \\
&- \frac{\hbar^2}{2m_B}
		\frac{\partial^2}{\partial z^2} + \frac{m_B\omega^2}{2} z^2 +G\sqrt{\frac{\hbar^3\omega}{m_B}}\left[\delta(z- x)+\sum_{i=1}^{2}\delta(z- y_i)\right]
\label{Ham3-1}\\
\hat{\cal H}_c &= \sqrt{\frac{\hbar^3\omega}{m_B}}\sum_{i=1}^{2}\delta(x - y_i). \label{Ham3-2}
\end{eqnarray} \end{numparts}
From the experimental perspective, this model can be viewed as a generalization of the previous model of $^{40}$K-$^{6}$Li mixture to cases when an additional $^{6}$Li atom in a different hyperfine state is present in the system. The parameter $G$ controls interactions between the additional particle with the two-component system. We aim to examine the performance of the control protocol for different values of $G$. Particularly, we want to answer the question if in the presence of interactions with a third-component particle, we are able to obtain the desired target state of two components by optimizing solely interactions within this subsystem. 

We assume that at the beginning the system is prepared in the non-interacting ground state, {\it i.e.}, it can be written as a product state $|\Psi(0)\rangle=|\mathtt{ini}\rangle|\Phi_0\rangle$, where $|\mathtt{ini}\rangle$ is the initial state of two-component mixture exactly as considered before while $|\Phi_0\rangle$ is the lowest single-particle orbital describing third component particle in its harmonic trap. The evolution of a whole system is governed by the Hamiltonian (6) and thus its quantum state $|\Psi(t)\rangle$ evolves according to the time-dependent Schr\"odinger equation. To have a full correspondence to the previously studied cases, we demand that at the final instant, $T$ the state of the system is as close as possible to the product state $|{\Psi(T)}\rangle = |\mathtt{tar}\rangle|\Phi(T)\rangle$. In this way we demand that the two-component subsystem is driven by optimized protocol to a desired target state $|\mathtt{tar}\rangle$,  the state of the third-component particle is arbitrary, and there are no correlations between these two subsystems. Since during the evolution, the state of the system is not necessarily a product state, we will calculate all temporal properties of the two-component subsystem from its reduced density matrix obtained by tracing out the third-component particle
\begin{equation}
\hat{\rho}(t) = \mathrm{Tr}_C\Bigl( |\Psi(t)\rangle\langle\Psi(t)|\Bigl).
\end{equation}
Particularly, the temporal fidelity with the target state (\ref{TempFid}) is now defined as
\begin{equation}
F(g(t),T)=\left[\mathrm{Tr}\Bigl(\hat{\rho}(t)|\mathtt{tar}\rangle\langle\mathtt{tar}| \Bigl)\right]^2 = |\langle\mathtt{tar}|\hat{\rho}(t)|\mathtt{tar}\rangle|^2.
\end{equation}
\begin{figure}
\includegraphics[width=\linewidth]{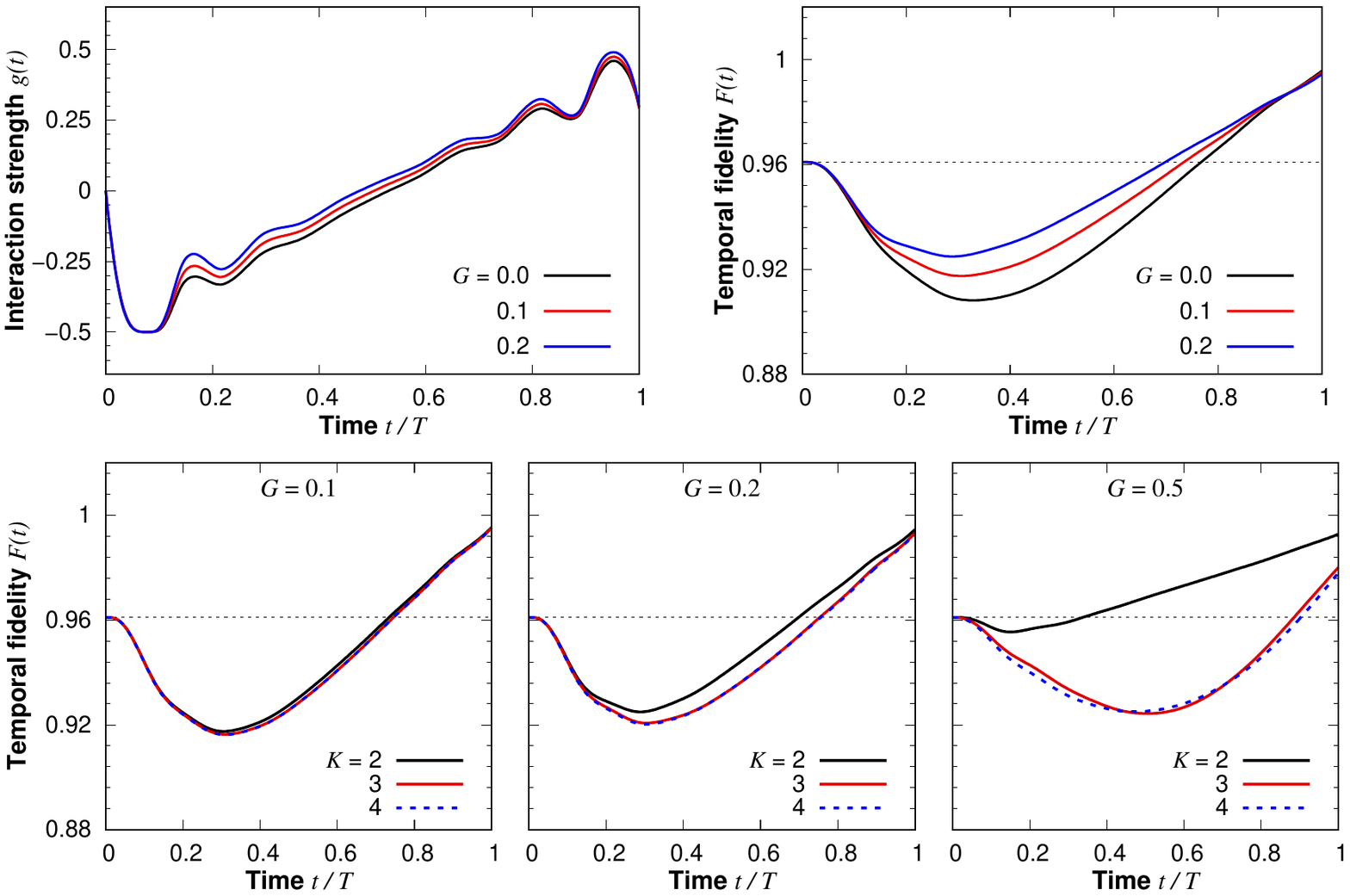}
	\caption{(top) Time evolution of interaction strength and corresponding temporal fidelity for the two-component three fermions system subjected to additional constant interaction $G$ with the third-component particle. Along with increasing interaction $G$ small modifications of optimized driving are needed to achieve the desired final fidelity. (bottom) Sensitivity of temporal fidelity $F(t)$ on the cut-off $K$ for different interaction strengths with the third component. For sufficiently low $G$ limiting the Hilbert space of the third-component particle to two the lowest orbitals is legitimized.}\label{Fig6}
\end{figure}
To get a better comparison with previously studied two-component scenarios, let us focus on the simplest generalization of a previous case with the target interaction $g_2=1$ and allowed range of control field $g\in [-0.5,0.5]$. We will check the performance of the optimization method for different strengths of interactions. In the top row of Fig.~\ref{Fig6} we show the results obtained for this scenario, assuming protocol duration $T=3\omega^{-1}$, for three different interactions $G$. In the left plot, we display optimized time dependence of the control field $g(t)$ while in the right plot the temporal fidelity $F(t)$ is presented. We notice that along with increasing interaction with the third component $G$ one needs to slightly modify the control field to maximize the final fidelity. Consequently, the temporal fidelity $F(t)$ is slightly modified, however the final fidelity is saturated close to the same value. It means that the final state (after integrating out the third-component particle) is very close to the desired target state $|\mathtt{tar}\rangle$. From this perspective, similarly, as it was checked for imperfections in the control field, we can argue that although interactions with the external system lead to some changes in the dynamics of the system, after careful treatment they are not very destructive for the performance of the proposed protocol.

All the calculations presented in the top row of Fig.~\ref{Fig6} were performed for relatively small interactions $G$. This allowed us to reduce the Hilbert space of the third component to $K=2$ single-particle orbitals (of course we keep cut-off $C=14$ for the first two components). To show that the results are indeed well-converged for these cut-offs, in the bottom row of Fig.~\ref{Fig6} we display temporal fidelities when the optimal protocol found for $K=2$ is applied to the system with higher $K$. We say that the cut-off $K=2$ is sufficient if the resulting fidelity is not affected by expanding the Hilbert space to a larger $K$. Taking such an approach as decisive we find that for $G=0.1$ the results are almost the same for larger cut-offs while for $G=0.2$ some small deviations in the middle moments are visible. However, the final fidelity is still almost the same. For stronger interactions (an example for $G=0.5$ is presented) the optimal protocol determined for $K=2$ gives clearly different results when applied to systems with higher $K$. Therefore, we conclude that the cut-off $K=2$ is insufficient in this case. Thus, in the top row, we limit ourselves only to sufficiently small interactions. Of course, in principle, the method presented can be straightforwardly extended to include stronger interactions with the third component or to increase the number of particles. This however would require much larger computational resources and is beyond the scope of this work.

\section{Conclusion}\label{sec:conclusion}
By using quantum optimal control, we have proposed a systematic and effective scheme for steering small quantum systems toward ground states that are inaccessible due to experimental limitations of the available range of the control parameter. As generic illustrations, we have demonstrated the viability of the scheme for model systems like an interacting two-qubit model or two- and three-component mixture of a few ultra-cold fermions. In the latter case, we considered a scenario for which the target state is demanded only for the two-component subsystem. In all these cases we demonstrated that the selected target state can be obtained with nearly perfect fidelity and with fairly finite duration. In addition, we have shown the robustness of the scheme by considering the perturbation to the optimized protocols. 

Our findings can be particularly important and useful for studying many-body systems that display exotic properties only for extreme values of control parameters. It also highlights the potential efficiency of optimal control techniques in advancing experimental realizations of quantum simulations and computations.

\section*{Data availability statement}
All numerical data presented in this paper is freely available online~\cite{Zenodo}.

\section*{Acknowledgements}
This research was supported by the Science Research Project of the Anhui Educational Committee (2023AH050073) and by the National Science Centre (NCN, Poland) within the OPUS project No. 2023/49/B/ST2/03744 (TS). For the purpose of Open Access, the authors have applied a CC-BY public copyright licence to any Author Accepted Manuscript version arising from this submission.

\section*{Bibliography}

\bibliographystyle{iopart-num}
\bibliography{main}

\providecommand{\newblock}{}
\begin{thebibliography}{10}
\expandafter\ifx\csname url\endcsname\relax
  \def\url#1{{\tt #1}}\fi
\expandafter\ifx\csname urlprefix\endcsname\relax\def\urlprefix{URL }\fi
\providecommand{\eprint}[2][]{\url{#2}}

\bibitem{2018AcinRoadMap}
Acín A, Bloch I, Buhrman H, Calarco T, Eichler C, Eisert J, Esteve D, Gisin N,
  Glaser S~J, Jelezko F, Kuhr S, Lewenstein M, Riedel M~F, Schmidt P~O, Thew R,
  Wallraff A, Walmsley I and Wilhelm F~K 2018 {\em New Journal of Physics\/}
  {\bf 20} 080201 \urlprefix\url{https://dx.doi.org/10.1088/1367-2630/aad1ea}

\bibitem{2014NoriRevModPhys}
Georgescu I~M, Ashhab S and Nori F 2014 {\em Rev. Mod. Phys.\/} {\bf 86}(1)
  153--185 \urlprefix\url{https://link.aps.org/doi/10.1103/RevModPhys.86.153}

\bibitem{2012BlochNatPhys}
Bloch I, Dalibard J and Nascimb{\`e}ne S 2012 {\em Nature Physics\/} {\bf 8}
  267--276 \urlprefix\url{https://doi.org/10.1038/nphys2259}

\bibitem{2014DevoretBook}
Devoret M, Huard B, Schoelkopf R and Cugliandolo L~F 2014 {\em Quantum
  Machines: Measurement and Control of Engineered Quantum Systems: Lecture
  Notes of the Les Houches Summer School: Volume 96, July 2011\/} (Oxford
  University Press) ISBN 9780199681181
  \urlprefix\url{https://doi.org/10.1093/acprof:oso/9780199681181.001.0001}

\bibitem{Glaser2015}
Glaser S~J, Boscain U, Calarco T, Koch C~P, K{\"o}ckenberger W, Kosloff R,
  Kuprov I, Luy B, Schirmer S, Schulte-Herbr{\"u}ggen T, Sugny D and Wilhelm
  F~K 2015 {\em The European Physical Journal D\/} {\bf 69} 279
  \urlprefix\url{https://doi.org/10.1140/epjd/e2015-60464-1}

\bibitem{2022KochEPJQ}
Koch C~P, Boscain U, Calarco T, Dirr G, Filipp S, Glaser S~J, Kosloff R,
  Montangero S, Schulte-Herbr{\"u}ggen T, Sugny D and Wilhelm F~K 2022 {\em EPJ
  Quantum Technology\/} {\bf 9} 19
  \urlprefix\url{https://doi.org/10.1140/epjqt/s40507-022-00138-x}

\bibitem{Wen2017}
Wen X~G 2017 {\em Rev. Mod. Phys.\/} {\bf 89}(4) 041004
  \urlprefix\url{https://link.aps.org/doi/10.1103/RevModPhys.89.041004}

\bibitem{2025GrassRMP}
Grass T, Bercioux D, Bhattacharya U, Lewenstein M, Nguyen H~S and Weitenberg C
  2025 {\em Rev. Mod. Phys.\/} {\bf 97}(1) 011001
  \urlprefix\url{https://link.aps.org/doi/10.1103/RevModPhys.97.011001}

\bibitem{Deffner2017}
Deffner S and Campbell S 2017 {\em Journal of Physics A: Mathematical and
  Theoretical\/} {\bf 50} 453001
  \urlprefix\url{https://dx.doi.org/10.1088/1751-8121/aa86c6}

\bibitem{dalessandro2021}
{D’Alessandro} D 2021 {\em Introduction to Quantum Control and Dynamics\/}
  (Chapman and Hall/CRC) \urlprefix\url{https://doi.org/10.1201/9781003051268}

\bibitem{Brif2010}
Brif C, Chakrabarti R and Rabitz H 2010 {\em New Journal of Physics\/} {\bf 12}
  075008 \urlprefix\url{https://dx.doi.org/10.1088/1367-2630/12/7/075008}

\bibitem{KHANEJA2005}
Khaneja N, Reiss T, Kehlet C, Schulte-Herbruggen T and Glaser S~J 2005 {\em
  Journal of Magnetic Resonance\/} {\bf 172} 296--305 ISSN 1090-7807
  \urlprefix\url{https://www.sciencedirect.com/science/article/pii/S1090780704003696}

\bibitem{vanFrank2016}
van Frank S, Bonneau M, Schmiedmayer J, Hild S, Gross C, Cheneau M, Bloch I,
  Pichler T, Negretti A, Calarco T and Montangero S 2016 {\em Scientific
  Reports\/} {\bf 6} 34187 ISSN 2045-2322
  \urlprefix\url{https://doi.org/10.1038/srep34187}

\bibitem{Li2018}
Li X, Pecak D, Sowi\ifmmode~\acute{n}\else \'{n}\fi{}ski T, Sherson J and
  Nielsen A~E~B 2018 {\em Phys. Rev. A\/} {\bf 97}(3) 033602
  \urlprefix\url{https://link.aps.org/doi/10.1103/PhysRevA.97.033602}

\bibitem{2019FogartySciPost}
Fogarty T, Ruks L, Li J and Busch T 2019 {\em SciPost Phys.\/} {\bf 6} 021
  \urlprefix\url{https://scipost.org/10.21468/SciPostPhys.6.2.021}

\bibitem{2019KahanUniv}
Kahan A, Fogarty T, Li J and Busch T 2019 {\em Universe\/} {\bf 5} ISSN
  2218-1997 \urlprefix\url{https://www.mdpi.com/2218-1997/5/10/207}

\bibitem{Sklarz2002}
Sklarz S~E and Tannor D~J 2002 {\em Phys. Rev. A\/} {\bf 66}(5) 053619
  \urlprefix\url{https://link.aps.org/doi/10.1103/PhysRevA.66.053619}

\bibitem{krotov1993}
Krotov V~F 1993 {\em Global Methods in Optimal Control Theory\/} (Boston, MA:
  Birkh{\"a}user Boston) pp 74--121 ISBN 978-1-4612-0349-0
  \urlprefix\url{https://doi.org/10.1007/978-1-4612-0349-0_3}

\bibitem{Doria2011}
Doria P, Calarco T and Montangero S 2011 {\em Phys. Rev. Lett.\/} {\bf 106}(19)
  190501
  \urlprefix\url{https://link.aps.org/doi/10.1103/PhysRevLett.106.190501}

\bibitem{Sorensen2018}
S\o{}rensen J~J~W~H, Aranburu M~O, Heinzel T and Sherson J~F 2018 {\em Phys.
  Rev. A\/} {\bf 98}(2) 022119
  \urlprefix\url{https://link.aps.org/doi/10.1103/PhysRevA.98.022119}

\bibitem{Machnes2018}
Machnes S, Ass\'emat E, Tannor D and Wilhelm F~K 2018 {\em Phys. Rev. Lett.\/}
  {\bf 120}(15) 150401
  \urlprefix\url{https://link.aps.org/doi/10.1103/PhysRevLett.120.150401}

\bibitem{Storn1997}
Storn R and Price K 1997 {\em Journal of Global Optimization\/} {\bf 11}
  341--359 ISSN 1573-2916
  \urlprefix\url{https://doi.org/10.1023/A:1008202821328}

\bibitem{das2011}
Das S and Suganthan P~N 2011 {\em IEEE Transactions on Evolutionary
  Computation\/} {\bf 15} 4--31

\bibitem{Hansen2006}
Hansen N 2006 {\em The CMA Evolution Strategy: A Comparing Review\/} (Berlin,
  Heidelberg: Springer Berlin Heidelberg) pp 75--102 ISBN 978-3-540-32494-2
  \urlprefix\url{https://doi.org/10.1007/3-540-32494-1_4}

\bibitem{Lloyd2014}
Lloyd S and Montangero S 2014 {\em Phys. Rev. Lett.\/} {\bf 113}(1) 010502
  \urlprefix\url{https://link.aps.org/doi/10.1103/PhysRevLett.113.010502}

\bibitem{Boscain2006}
Boscain U and Mason P 2006 {\em Journal of Mathematical Physics\/} {\bf 47}
  062101 ISSN 0022-2488 (\textit{Preprint}
  \eprint{https://pubs.aip.org/aip/jmp/article-pdf/doi/10.1063/1.2203236/13949066/062101\_1\_online.pdf})
  \urlprefix\url{https://doi.org/10.1063/1.2203236}

\bibitem{Hegerfeldt2013}
Hegerfeldt G~C 2013 {\em Phys. Rev. Lett.\/} {\bf 111}(26) 260501
  \urlprefix\url{https://link.aps.org/doi/10.1103/PhysRevLett.111.260501}

\bibitem{Jafarizadeh2020}
Jafarizadeh M, Naghdi F and Bazrafkan M 2020 {\em Physics Letters A\/} {\bf
  384} 126743 ISSN 0375-9601
  \urlprefix\url{https://www.sciencedirect.com/science/article/pii/S0375960120306101}

\bibitem{chong2013}
Edwin K P~Chong S~H~Z 2013 {\em An Introduction to Optimization\/} (John Wiley
  \& Sons)
  \urlprefix\url{https://www.wiley.com/en-ie/An+Introduction+to+Optimization%2C+4th+Edition-p-9781118515150}

\bibitem{Bukov2018}
Bukov M, Day A~G~R, Weinberg P, Polkovnikov A, Mehta P and Sels D 2018 {\em
  Phys. Rev. A\/} {\bf 97}(5) 052114
  \urlprefix\url{https://link.aps.org/doi/10.1103/PhysRevA.97.052114}

\bibitem{2018LiewPRB}
Liew T~C~H and Rubo Y~G 2018 {\em Phys. Rev. B\/} {\bf 97}(4) 041302
  \urlprefix\url{https://link.aps.org/doi/10.1103/PhysRevB.97.041302}

\bibitem{2018StefanatosPRB}
Stefanatos D and Paspalakis E 2018 {\em Phys. Rev. B\/} {\bf 98}(3) 035303
  \urlprefix\url{https://link.aps.org/doi/10.1103/PhysRevB.98.035303}

\bibitem{2020StefanatosPRA}
Stefanatos D and Paspalakis E 2020 {\em Phys. Rev. A\/} {\bf 102}(1) 013716
  \urlprefix\url{https://link.aps.org/doi/10.1103/PhysRevA.102.013716}

\bibitem{Frey2016}
Frey M~R 2016 {\em Quantum Information Processing\/} {\bf 15} 3919--3950 ISSN
  1573-1332 \urlprefix\url{https://doi.org/10.1007/s11128-016-1405-x}

\bibitem{2009CanevaPRL}
Caneva T, Murphy M, Calarco T, Fazio R, Montangero S, Giovannetti V and Santoro
  G~E 2009 {\em Phys. Rev. Lett.\/} {\bf 103}(24) 240501
  \urlprefix\url{https://link.aps.org/doi/10.1103/PhysRevLett.103.240501}

\bibitem{2022HornedalNJP}
Hörnedal N, Allan D and Sönnerborn O 2022 {\em New Journal of Physics\/} {\bf
  24} 055004 \urlprefix\url{https://dx.doi.org/10.1088/1367-2630/ac688a}

\bibitem{2024VikramPRL}
Vikram A and Galitski V 2024 {\em Phys. Rev. Lett.\/} {\bf 132}(4) 040402
  \urlprefix\url{https://link.aps.org/doi/10.1103/PhysRevLett.132.040402}

\bibitem{2012BlumeRPP}
Blume D 2012 {\em Reports on Progress in Physics\/} {\bf 75} 046401
  \urlprefix\url{https://dx.doi.org/10.1088/0034-4885/75/4/046401}

\bibitem{2019SowinskiRPP}
Sowiński T and Ángel García-March M 2019 {\em Reports on Progress in
  Physics\/} {\bf 82} 104401
  \urlprefix\url{https://dx.doi.org/10.1088/1361-6633/ab3a80}

\bibitem{2023MistakidisPhysRep}
Mistakidis S, Volosniev A, Barfknecht R, Fogarty T, Busch T, Foerster A,
  Schmelcher P and Zinner N 2023 {\em Physics Reports\/} {\bf 1042} 1--108 ISSN
  0370-1573 few-body Bose gases in low dimensions—A laboratory for quantum
  dynamics
  \urlprefix\url{https://www.sciencedirect.com/science/article/pii/S0370157323003162}

\bibitem{Zenodo}
Li X and Sowi\'nski T 2025 {Data for "Swinging small quantum systems out of
  available values of control parameters"} {Zenodo}
  \urlprefix\url{https://doi.org/10.5281/zenodo.15223013}

\end{thebibliography}

\end{document}